\documentclass[%
 superscriptaddress,
 preprint,
 showpacs,preprintnumbers,
 amsmath,amssymb,
 aps,
 prb,
]{revtex4-2}

\usepackage{graphicx}
\usepackage{dcolumn}
\usepackage{bm}

\usepackage{upgreek}
\usepackage{amsmath}
\begin{document}


\title{Enhanced second harmonic generation in high-$Q$ all-dielectric metasurfaces with backward frequency conversion}

\author{Xu Tu}
\affiliation{School of Information Engineering, Nanchang University, Nanchang 330031, China}
\affiliation{Institute for Advanced Study, Nanchang University, Nanchang 330031, China}

\author{Siqi Feng}
\affiliation{School of Information Engineering, Nanchang University, Nanchang 330031, China}
\affiliation{Institute for Advanced Study, Nanchang University, Nanchang 330031, China}

\author{Jiajun Li}
\affiliation{School of Information Engineering, Nanchang University, Nanchang 330031, China}
\affiliation{Institute for Advanced Study, Nanchang University, Nanchang 330031, China}

\author{Yangguang Xing}
\affiliation{School of Information Engineering, Nanchang University, Nanchang 330031, China}
\affiliation{Institute for Advanced Study, Nanchang University, Nanchang 330031, China}

\author{Feng Wu}
\affiliation{School of Optoelectronic Engineering, Guangdong Polytechnic Normal University, Guangzhou 510665, China}

\author{Tingting Liu}
\email{ttliu@ncu.edu.cn}
\affiliation{School of Information Engineering, Nanchang University, Nanchang 330031, China}
\affiliation{Institute for Advanced Study, Nanchang University, Nanchang 330031, China}

\author{Shuyuan Xiao}
\email{syxiao@ncu.edu.cn}
\affiliation{School of Information Engineering, Nanchang University, Nanchang 330031, China}
\affiliation{Institute for Advanced Study, Nanchang University, Nanchang 330031, China}

\begin{abstract}
	
Here we employ the quasi-bound state in the continuum (quasi-BIC) resonance in all-dielectric metasurfaces for efficient nonlinear processes in consideration of the backward frequency conversion. We theoretically study the second-harmonic generation (SHG) from symmetry-broken AlGaAs metasurfaces and reveal the efficiency enhancement empowered by high-$Q$ quasi-BIC resonances. By introducing the correction term of nonlinear polarization at the fundamental wave field to the conventional undepleted approximation, we uncover the effect of backward frequency conversion on the nonlinear conversation efficiency. The SHG efficiency as $2.45\times10^{-2}$ with the developed depleted model, shows a $14.3\%$ decrease compared with $2.86\times10^{-2}$ in conventional undepleted approximation, under the incident intensity of 10 MW/cm$^{2}$. Our results are of significant importance for designing efficient nonlinear metasurfaces supporting high-$Q$ resonances.

\end{abstract}

\maketitle


\section{\label{sec1}Introduction}

Frequency conversion processes, which involve the interaction between intense light and matter, have made consideration advancements following the development of laser\cite{Boyd2008}. Harmonic generation processes such as second- (SHG) and third-harmonic generation (THG) are the most fundamental physical phenomena in nonlinear optics, and have attracted growing interest in a large variety of optical devices. They are traditional realized in the bulk nonlinear crystals, where the cumbersome phase matching condition is required to satisfy momentum match for efficient frequency conversion. The typical bulk size and the phase match constraint are incompatible with advanced technologies. In recent years, advances in nanofabrication techniques have revolutionized the framework of nonlinear processes. Ultrathin metasurfaces are of great promise for replacing the bulk crystals since they are capable to relax the constraint of phase matching and simultaneously manipulate nonlinear fields with favorable conversion efficiecies\cite{Li2017,Krasnok2018}. For example, plasmonic metasurfaces have been firstly used in harmonic generations, frequency mixing, and other nonlinear effects\cite{Kauranen2012,Panoiu2018}. Since the plasmonic mechanism relies on the collective oscillation of free electrons, the inevitable ohmic loss and heat effect would lead to low optical damage threshold. These shortcomings in metallic nano-resonators inherently limits the nonlinear conversion efficiency and then hinders the applicability in nonlinear nanophotonic devices.

Recently, all-dielectric metasurfaces have been demonstrated to enhance and manipulate nonlinear frequency conversation processes at the nanoscale\cite{Sain2019,Grinblat2021,Liu2022}. In addition to the advantages of low loss, high refractive index, and high damage threshold, they exhibit the multipolar characteristics with the coexistence of electric and magnetic resonant modes known as Mie-type resonances, giving rise to the strong localization of electromagnetic fields beneficial for the high nonlinear conversation efficiency. In particular, bound states in the continuum (BICs) in all-dielectric metasurfaces have been proved to be a promising way to achieve higher efficiencies of nonlinear optical processes by engineering the radiation leakage. When the radiation channel is opened, sharp quasi-BIC resonances with high quality factors ($Q$ factors) are formed accompanied by the strong local field enhancement\cite{Hsu2016,Huang2023}. It was reported that such high-$Q$ quasi-BIC resonances can boost substantially nonlinear frequency conversion processes in dielectric metasurfaces such as SHG\cite{Vabishchevich2018,Carletti2018,Ning2021,Cai2023}, THG\cite{Liu2019,Xu2019,Koshelev2019}, high-harmonic generation (HHG)\cite{Carletti2019,Zograf2022,Xiao2022,Zalogina2023}, sum-frequency generation (SFG)\cite{CamachoMorales2022, Zhang2022, Feng2023, Liu2024}, and four-wave mixing (FWM)\cite{Xu2022, Liu2023, Sanderson2024, Moretti2024}. Indeed, we notice the the quest for high conversion efficiency is rooted in the enhancement of local field inside the nonlinear nanostructures. Conventionally, the nonlinear frequency conversion process is simply attributed to the forward frequency conversion of the fundamental waves to nonlinear waves. However, when the local field is large enough, the influence of backward coupling, i.e., the backward frequency conversion of the nonlinear waves to the fundamental waves could not be negligible. 

In this work, we study the efficient nonlinear SHG process in quasi-BIC dielectric metasurfaces in consideration of backward frequency conversion. We apply the approach as a depleted model to nonlinear AlGaAs metasurfaces where the excitation of the quasi-BIC resonance allows for the extremely enhanced local fields, leading to the high efficiency of SHG process. We show that the conversion efficiency of SHG becomes lower in quasi-BIC dielectric metasurfaces, when the backward frequency conversion is taken into account in the form of the correction term of nonlinear polarization at the fundamental wave field. In the designed metasurface, consideration of the backward frequency conversion combined with the forward frequency conversion is necessary with the pump intensity of 10 MW/cm$^{2}$ where the difference of SHG efficiency calculated based on the undepleted and depleted models approaches $14.3\%$.

\section{\label{sec2}High-$Q$ quasi-BIC resonance in AlGaAs metasurfaces}

\begin{figure*}[htbp]
\centering
\includegraphics
[scale=0.4]{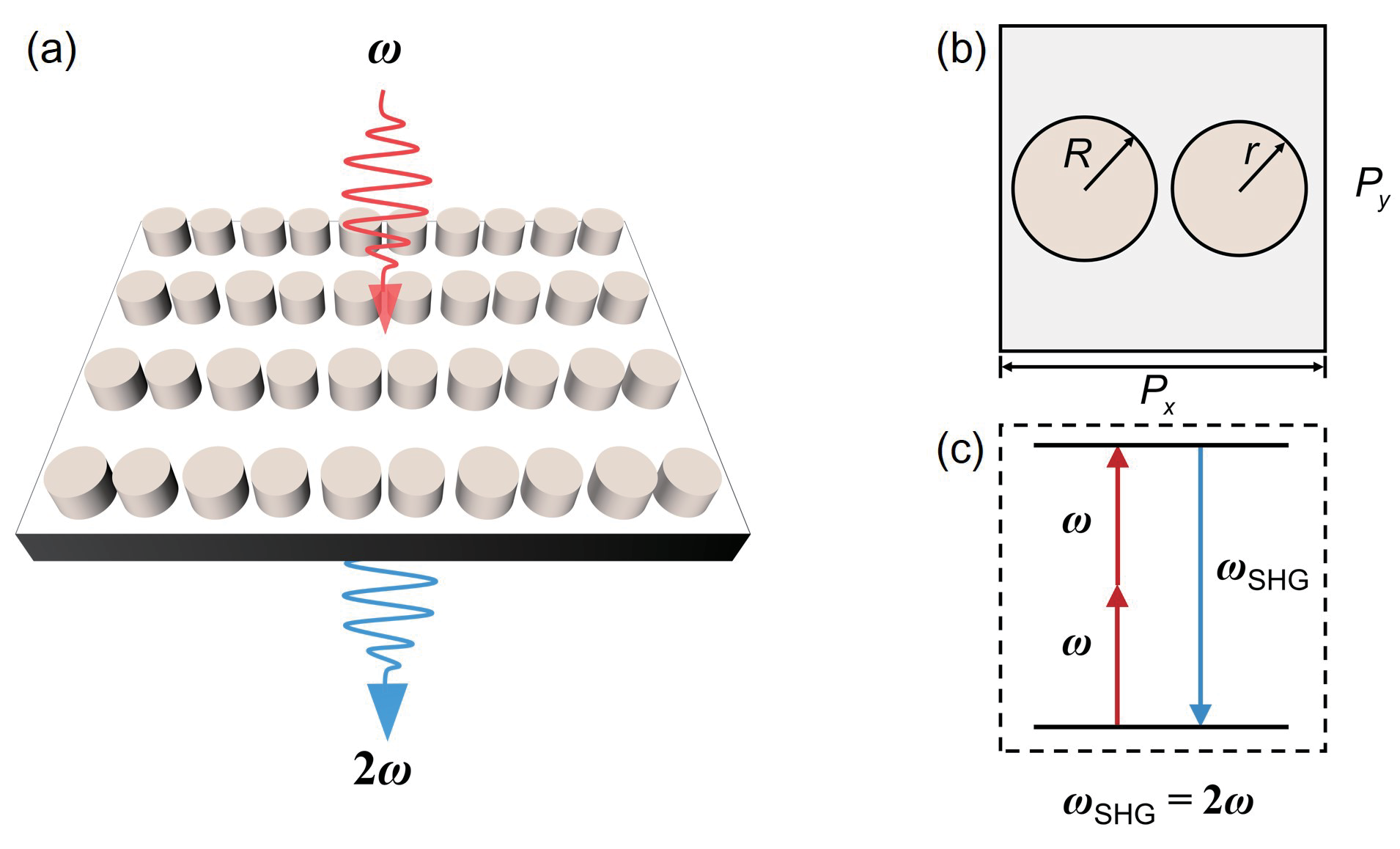}
\caption{\label{fig1} (a) The conceptual schematic of SHG process in the quasi-BIC metasurfaces. (b) The planar portion of the unit cell. Geometric parameters: the period is $p_{x}=p_{y}=1020$ nm, the height is $h=300$ nm, and the radii are initially set to $R=r=225$ nm. (c) The energy diagram of the SHG process. }
\end{figure*}

We consider the resonant AlGaAs metasurfaces for nonlinear SHG process, as shown in Fig. \ref{fig1}. The fundamental pump wave is normally incident to the metasurface structure, and the transmitted SHG signals are collected from the substrate side. AlGaAs is chosen because of its strong second-order nonlinear susceptibility for SHG response, and its high refractive index that is necessary to sustain the quasi-BIC resonances in the wavelength of interest\cite{CamachoMorales2016,Rocco2024}. The metasurfaces are composed of a square lattice of patterned AlGaAs nanodisks of height $h=300$ nm on top of a SiO$_{2}$ substrate with the period of $p_{x}=p_{y}=1020$ nm. The refractive index of AlGaAs is derived from experimental data and plotted in Fig. S1\cite{SM} (see also Ref.\cite{Gehrsitz2000} therein), and the refractive index of the substrate is set to 1.45. The radii of circular-circular nanodisks are $R=r=225$ nm. A tunable geometrical parameter here is the radius $r$ of the nanodisk at the right side.

\begin{figure*}[htbp]
\centering
\includegraphics
[scale=0.4]{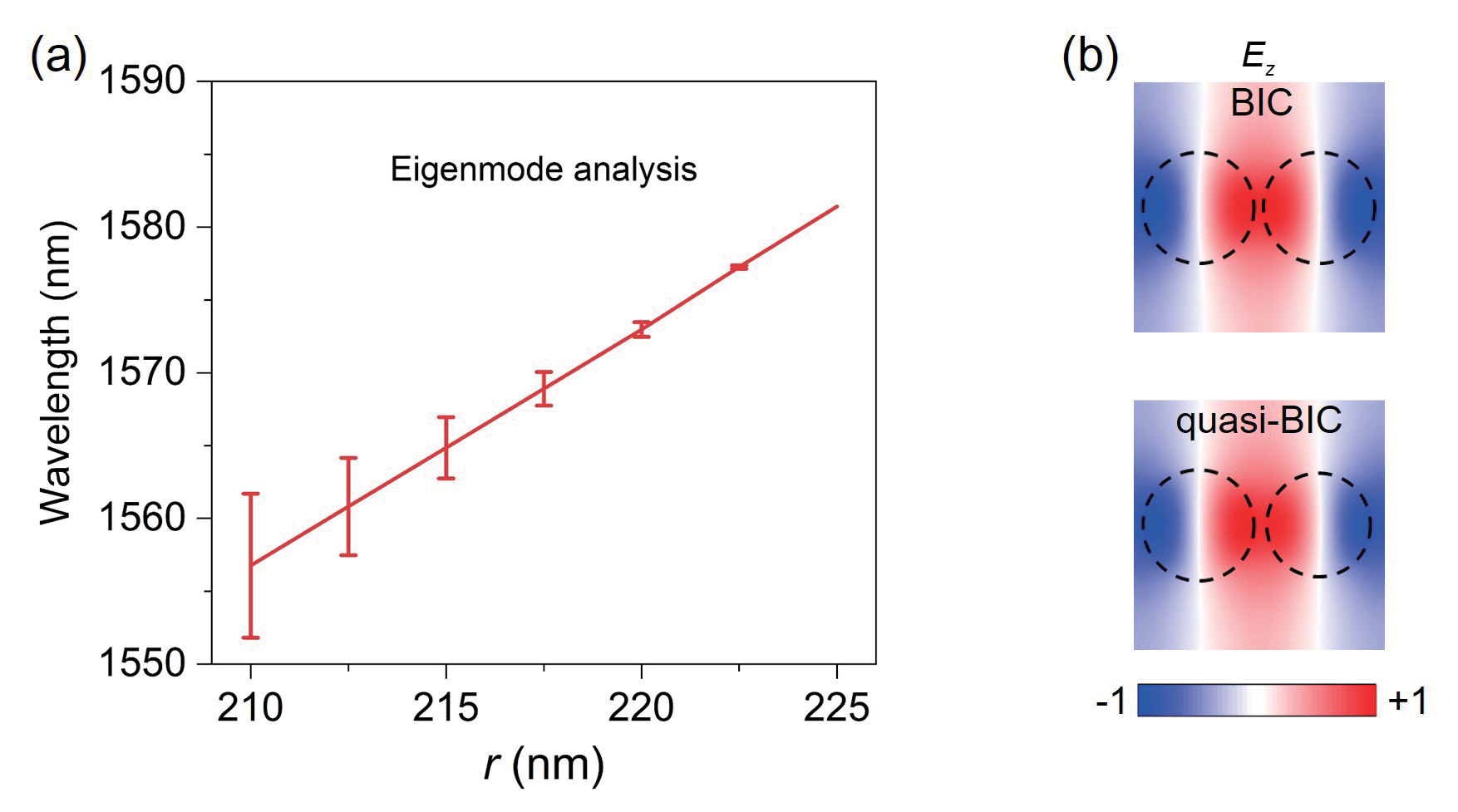}
\caption{\label{fig2} (a) The eigenmode of the quasi-BIC metasurfaces as a function of the pump wavelength and the radius $r$. The error bars represent the relative magnitude of the mode inverse radiation lifetime. (b) The distribution of magnetic fields in the $z$-direction $E_{z}$ for both BIC and quasi-BIC.}
\end{figure*}

We start from the eigenmode analysis to calculate and optimize the resonant mode of the metasurface system using the finite element method via COMSOL Multiphysics. The metasurfaces are considered as an infinite structure with perfect periodicity. Fig. \ref{fig2}(a) shows the resonant modes of the structure as a function of the geometrical parameter of the radius $r$. The error bars represent the relative magnitude of the mode inverse radiation lifetime, i.e. radiation loss $\gamma$. Here we would like to use the relative length of the error bars to more intuitively characterize the increase of BIC eigenmode inverse radiation lifetime with the structural symmetry breaking. In other words, we do not focus on the specific unit but rather on the relative length of the error bars. The lengths of error bars are normalized via multiplying them by a constant coefficient. The values of eigenfrequency including the real and imaginary parts are listed in Table S1\cite{SM}. It is observed that the variations of $r$ change the eigenfrequency and radiation loss of the system. At $r=R=225$ nm, the metasurfaces support a symmetry-protected BIC that is completely decoupled from the external free space as a non-radiative mode with $\gamma=0$. To transform the symmetry-protected BIC into a leaky mode, we break the structural symmetry by changing $r$ to provide a certain radiation leakage channel. Thus, as $r$ deviates from 225 nm, the genuine BIC is transformed into the quasi-BIC with increasing radiation loss. This evolution can also be validated by their electric field distributions $E_{z}$ in Fig. \ref{fig2}(b) where the in-plane inversion symmetry $(x, y)\rightarrow(-x,-y)$ of $E_{z}$ at BIC is slightly broken by the structural perturbation. 

\begin{figure*}[htbp]
\centering
\includegraphics
[scale=0.4]{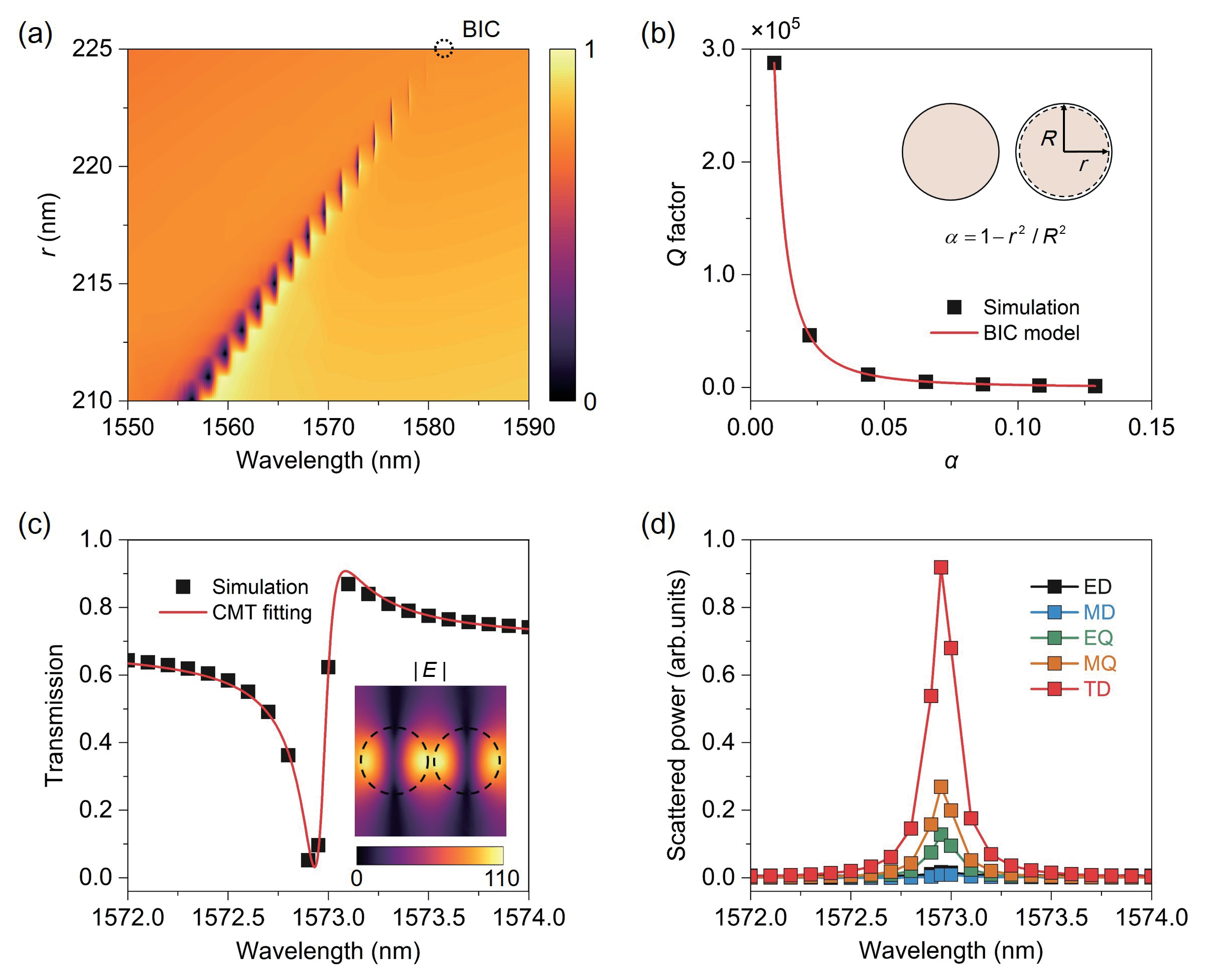}
\caption{\label{fig3} (a) The transmission spectra of the quasi-BIC metasurfaces as a function of wavelength and radius $r$. (b) The quadratic dependence of $Q$ factor on the asymmetry parameter $\alpha$. (c) The simulated and theoretically fitted transmission spectrum for the quasi-BIC metasurface with $r=220$ nm. The inset shows the distribution of the electric field in the $x$-$y$ plane at the corresponding resonant wavelength. (d) The multipolar decomposition of scattered power for the metasurface with $r=220$ nm.}
\end{figure*}

We proceed to the linear response of the metasurfaces under the normally incident $x$-polarized plane wave. Figure \ref{fig3}(a) depicts the transmission spectra with varying radius $r$. When the structure is symmetric with $r=R=225$ nm, the transmission spectra display a disappearing dip that broadens as $r$ becomes smaller. This reveals the BIC with infinite $Q$ factor transforms into finite-$Q$ quasi-BIC, which confirms the results of the eigenmode analysis. As shown in Fig. \ref{fig3}(b), the $Q$ factor exhibits quadratic dependence on the asymmetric parameter $\alpha$ defined $\alpha=1-r^{2}/R^{2}$, i.e., $Q \propto \alpha^{-2}$. Such universal behavior for the $Q$ factor of a quasi-BIC remains valid for the proposed symmetry-broken metasurfaces\cite{Koshelev2018,Gao2022,Ma2023}. We provide the transmission spectrum for the metasurfaces with $r=220$ nm in Fig. \ref{fig3}(c). The asymmetric Fano shape is observed with the resonant wavelength at 1572.95 nm. According to the coupled-mode theory (CMT)\cite{Xu2019,Xiao2020,Wang2020}, the $Q$ factor can be calculated by the $\omega_\text{0}/2\gamma$ as 8950, indicating the remarkably confined field of the leaky quasi-BIC, with the strong field enhancement by a factor 110. Furthermore, the multipolar decomposition results in Fig. \ref{fig3}(d) reveal that the toroidal dipole (TD) plays the dominate role to this quasi-BIC resonance. The expressions of multipolar decomposition can be found in the Supplemental Material\cite{SM}, and see also Refs. \cite{Kaelberer2010,Savinov2014} therein.

\section{\label{sec3}Resonantly enhanced SHG with backward frequency conversion}

We further come to the quasi-BIC enhanced nonlinear response of the metasurfaces. The coupling wave equation at the fundamental and harmonic wavelength is written as\cite{saleh2019fundamentals}
\begin{equation}
	\nabla^{2}\vec{E}-\frac{1}{c^{2}}\frac{\partial^{2}\vec{E}}{\partial t^{2}}=\mu_{0}\frac{\partial^{2}\vec{P}_\text{nl}}{\partial t^{2}},
\label{equation1}
\end{equation}
where $c=c_{0}/n$, $n=1+\chi$, $c_{0}=1/\sqrt{\varepsilon_{0}\mu_{0}}$ is the speed of light in vacuum, $\chi$ is the linear polarizability, $\vec{E}$ is the electric field generated by the nonlinear polarization source, $P_\text{nl}$ is the nonlinear polarization. When only the forward frequency conversion is considered, $P_\text{nl}$ can be conventionally represented by the nonlinear polarization at the SH wavelength $P_\text{nl}^{(2\omega)}$. For the non-centrosymmetric materials such as AlGaAs adopted in this, the induced nonlinear polarization for SHG process is
\begin{equation}
	\vec{P}^{(2\omega)}=\varepsilon_{0}\chi^{(2)}\vec{E}^{(\omega)}\vec{E}^{(\omega)},
\label{equation2}
\end{equation}
where $\varepsilon_\text{0}$ is the vacuum dielectric constant with $8.8542\times10^{-12}$ F/m, $\chi^{(2)}$ is the second-order polarizability electric field inside the material. For AlGaAs material, only off-diagonal elements of the tensor $\chi^{2}$ is nonzero, i.e. $\chi_{ijk}^{(2)}=\chi^{(2)}=100$ pm/V, ${i}\neq{j}\neq{k}$. Therefore, Eq. (\ref{equation2}) can be explicitly written as
\begin{equation}
	\begin{bmatrix}
		P_{x}^{(2\omega)}\\
		P_{y}^{(2\omega)}\\
		P_{z}^{(2\omega)}
	\end{bmatrix}
	=2\varepsilon_{0}\chi^{(2)}
	\begin{bmatrix}
		E_{y}^{(\omega)}E_{z}^{(\omega)}\\
		E_{x}^{(\omega)}E_{z}^{(\omega)}\\
		E_{x}^{(\omega)}E_{y}^{(\omega)}
    \end{bmatrix}.
\label{equation3}
\end{equation}
Based on Eq. (\ref{equation3}), only the forward frequency conversion is considered. The nonlinear simulations of SHG process are implemented with the conventional undepleted pump approximation using two typical steps. Firstly, the linear response at the fundamental wavelength is calculated for the local field distributions, and the nonlinear polarization inside the meta-atom is induced. Secondly, the polarization term is employed as the only source to excite the electromagnetic field at the harmonic wavelength to generate SH radiation.

In the case of highly efficient frequency conversion, the electric field amplitudes inside the nonlinear meta-atoms at the fundamental and SH wavelengths may be comparable. When considering SHG as a degenerate three-wave mixing process, the forward frequency conversion $(\omega + \omega \rightarrow 2\omega)$ process and the backward frequency conversion $(2\omega + \omega^{*} \rightarrow \omega)$ processes are equally important in frequency conversion. Instead of the above undepleted approximation, the depleted model taking into account of forward and backward frequency conversion should be exploited in this case. Since the coupling of the nonlinear waves to the fundamental waves could not be neglected, $P_\text{nl}$ should include the nonlinear polarization at the SH wavelength $P_\text{nl}^{(2\omega)}$ and the correction term of nonlinear polarization at the fundamental wavelength $P_\text{nl}^{(\omega)}$ i.e., $P_\text{nl}=P_\text{nl}^{(\omega)}+P_\text{nl}^{(2\omega)}$. In the principle crystalline axis system such as the AlGaAs material, the nonlinear source $P_\text{nl}^{(2\omega)}$ at the SH wavelength is given by Eq. (\ref{equation3}), and expression for $P_\text{nl}^{(\omega)}$ at the fundamental wavelength is of the form\cite{Volkovskaya2020},
\begin{equation}
	\begin{bmatrix}
		P_{x}^{(\omega)}\\
		P_{y}^{(\omega)}\\
		P_{z}^{(\omega)}
	\end{bmatrix}
	=2\varepsilon_\text{0}\chi^{(2)}
	\begin{bmatrix}
		E_{y}^{(2\omega)}E_{z}^{(\omega)*}+E_{z}^{(2\omega)}E_{y}^{(\omega)*}\\
		E_{x}^{(2\omega)}E_{z}^{(\omega)*}+E_{z}^{(2\omega)}E_{x}^{(\omega)*}\\
		E_{x}^{(2\omega)}E_{y}^{(\omega)*}+E_{y}^{(2\omega)}E_{x}^{(\omega)*}
    \end{bmatrix}.
\label{equation4}
\end{equation}
In the depleted approximation model, the simulation for nonlinear SHG process follows two steps. After the first step to compute the linear response at the fundamental wavelength, the second step for nonlinear response considers both the nonlinear polarizations $P_\text{nl}^{(\omega)}$ and $P_\text{nl}^{(2\omega)}$ as sources in the fundamental and second harmonic frequencies, respectively, to excite the electromagnetic response that radiates the SHG signal. More details about the modeling methodology of nonlinear simulations can be found in the Supplemental Material\cite{SM}.

\begin{figure*}[htbp]
\centering
\includegraphics
[scale=0.4]{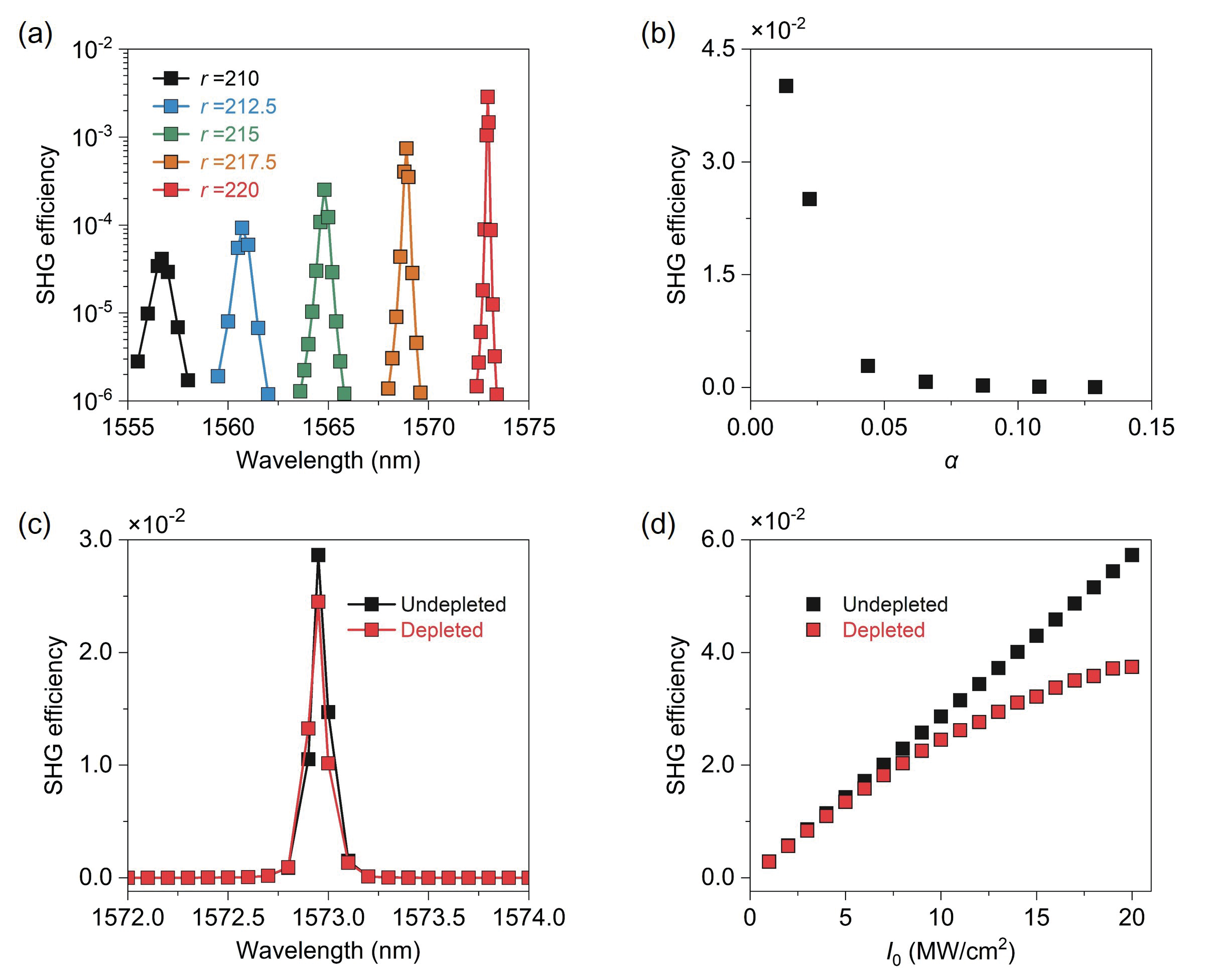}
\caption{\label{fig4} (a) The SHG conversion efficiencies with undepleted approximation in the quasi-BIC metasurfaces with the radius ranging from $r=210$ nm to 220 nm under incident pump intensity of 1 MW/cm$^{2}$. (b) The dependence of SHG efficiency on the asymmetry parameter $\alpha$. (c) The comparison of the SHG efficiencies with undepleted and depleted approximations in the quasi-BIC metasurface with $r=220$ nm under incident pump intensity of 10 MW/cm$^{2}$. (d) The SHG efficiencies with undepleted and depleted approximations as a function of the incident pump intensity.}
\end{figure*}

We first calculate the SHG efficiency with the undepleted pump approximation, under different asymmetric geometrical parameters of the metasurface. In simulations the incident light intensity is set to 1 MW/cm$^{2}$. The conversion efficiency of SHG is defined by $\eta_\text{SH}=P_\text{SH}/P_\text{FF}$, where $P_\text{SH}$ is the radiative power of the transmitted SH signal collected from the substrate side, $P_\text{FF}$ is the incident pump power at the fundamental wavelength. In Fig. \ref{fig4}(a), the conversion efficiency of SHG $\eta_\text{SH}$ increases with the increase of the radius $r$ in the quasi-BIC metasurface. For the proposed metasurface with $r=220$ nm, SHG efficiency is comparable or even further improved to previously theoretical results with quasi-BIC enhanced second-order nonlinear processes\cite{Carletti2018,Rocco2020,Yang2020,Ning2021}. Given that the geometrical asymmetry is related to the $Q$ factor of quasi-BIC resonance and the field confinement performance, we further explore the dependence of the SHG efficiency on the asymmetric parameters $\alpha$. It is observed in Fig. \ref{fig4}(b) that $\eta_\text{SH}$ is subject to $\alpha$, which is consistent with the quadratic dependence of $Q$ factor on $\alpha$. As asymmetric parameter increases, the radiation loss of the system increases, leading to the dramatical decline of $Q$ factor and electric field energy. As a result, the conversion efficiency of SHG process significantly decreases. 

We further take into consideration of the backward frequency conversion of the SH radiation field on the fundamental wave field. The SHG conversion efficiency is calculated with the depleted approximation in the quasi-BIC metasurface with $r=220$ nm. To clearly observe the effect of the backward frequency conversion, the incident pump is increased to 10 MW/cm$^{2}$. With the undepleted and depleted approximation, the SHG efficiencies are compared in Fig. \ref{fig4}(c). Both the efficiency curves display significant enhancement at the resonant wavelength owing to the high-$Q$ quasi-BIC resonance. The peak efficiency with undepleted approximation is calculated as $2.86\times10^{-2}$, higher than that with depleted model of $2.45\times10^{-2}$. The undepleted approximation causes error above $14.3\%$ in nonlinear simulations. This means that the system no longer completely satisfies the conditions for undepleted approximation due to the high enough SHG efficiency for SH field comparable with fundamental field. It is estimated that the nonlinear polarization $P_\text{nl}^{(\omega)}$ at the fundamental frequency cannot be ignored at high pump intensity.

In Fig. \ref{fig4}(d), we finally explore the dependence of the SHG efficiency on the incident pump intensity. The SHG conversion efficiency with undepleted approximation is proportional to the light intensity. This implies the nonlinear SHG efficiency can be enhanced by orders of magnitude by increasing the input light intensity. However, it is not feasible in practice when the backward frequency conversion is taken into account. With the backward frequency conversion of the SH radiation field on the fundamental field, the SHG efficiency would not increase linearly, but reach saturation as the pump intensity further increases, as shown in Fig. \ref{fig4}(d). In the proposed metasurface, the SHG efficiency is $3.74\times10^{-2}$ at pump intensity of 20 MW/cm$^{2}$ when the backward frequency conversion is considered, much smaller than the efficiency of $5.73\times10^{-2}$ with the undepleted approximation. This backward frequency conversion would be observable when implementing very high pump intensity\cite{Zalogina2023,Zograf2022,Gao2018,Timpu2019}. It is noticed that the irreversible damage of the nonlinear material AlGaAs may occur at very high incident pump intensity due to the increased absorption enhanced by the resonantly enhanced local electric field. Previous experiments have reported the damage threshold for the incident peak intensity is approximately several GW/cm$^{2}$. Specifically, the damage threshold $I_{\text{d}}$ is determined by two factors: 1) the intensity of the incident pump $I$; 2) the enhancement of the local electric field $f$, which can be expressed by a simple relation $I_{\text{d}}=f^{2}I$. Considering the experimentally estimated values of $f_{1}^{2}=30$ and $I_{1}=8.1$ GW/cm$^{2}$\cite{Liu2016,Gili2016}, and the theoretically calculated $f_{2}=110$ that can be observed from the inset in Fig. 3(c), the threshold of incident intensity can be obtained as $I_{2}=20.1$ MW/cm$^{2}$. Therefore, we would end with a maximum incident pump intensity $<20$ MW/cm$^{2}$ in Fig. 4(d). 

\section{\label{sec4}Conclusions}

In conclusion, we demonstrate the efficient nonlinear SHG process in the quasi-BIC dielectric metasurfaces. For the proposed engineered asymmetric AlGaAs metasurfaces, the SHG conversion efficiency can be enhanced up to be $10^{-2}$ order of magnitude under the incident intensity of 10 MW/cm$^{2}$. When taking the backward frequency conversion into consideration by introducing the correction term of nonlinear polarization at the fundamental wave field, we find that the conversion efficiency calculated by the developed depleted model becomes lower with around $14.3\%$ decrease compared with that with conventional undepleted approximation. Our calculation results are based on the general coupled wave equation through considering SHG as a degenerate three-wave mixing process, and would be more accordance with actual circumstance. The similar approach can be applied to the case of the SHG from other III-V semiconductor metasurfaces, where either ultrahigh-$Q$ resonances or sufficiently strong pump power is exploited. Thus, our general approach is of significant importance for designing efficient nonlinear metasurfaces supporting high-$Q$ resonances toward high-efficiency frequency conversion, optical switching, and modulation.

\begin{acknowledgments}	
This work was supported by the National Natural Science Foundation of China (Grants No. 12304420, 12264028, 12364045, and 12104105), the Natural Science Foundation of Jiangxi Province (Grants No. 20232BAB201040 and 20232BAB211025), the Guangdong Basic and Applied Basic Research Foundation (Grant No. 2023A1515011024), the Young Elite Scientists Sponsorship Program by JXAST (Grant No. 2023QT11), and the Innovation Fund for Graduate Students of Jiangxi Province (Grant No. YC2023-S028).	

The authors would also like to thank Dr. Lei Xu (Nottingham Trent University) for his guidance on the nonlinear simulations.
\end{acknowledgments}


%

\end{document}